\documentstyle[prl,twocolumn,aps]{revtex}
\def\new{\newcommand}
\def\renew{\renewcommand}
\new{\eq}[1]{(\ref{#1})}

\newtheorem{prop}{Proposition}
\new{\bd}{\begin{displaymath}}
\new{\ed}{\end{displaymath}}
\new{\be}{\begin{equation}}
\new{\ee}{\end{equation}}
\new{\bi}{\begin{itemize}}
\new{\ei}{\end{itemize}}
\new{\ba}{\begin{eqnarray}}
\new{\ea}{\end{eqnarray}}
\new{\bo}{\boldmath}
\new{\un}{\unboldmath}
\new{\mx}{\mbox}
\new{\bold}[1]{\mbox{\boldmath$#1$\unboldmath}}
\new{\bqu}{\begin{quote}}
\new{\equ}{\end{quote}}
\new{\sbs}[2]{\mbox{${#1}_{#2}$}}
\new{\subvel}[3]{\mbox{$#1_{#2},\ldots,#1_{#3}$}}
\new{\subvec}[3]{\mbox{$#1_{#2},\cdots,#1_{#3}$}}
\new{\avec}[2]{\mbox{${#1}_{1},\ldots,{#1}_{#2}$}}
\new{\spin}[2]{ \mbox{$ \bold{\sigma}_{#1} {\bf \cdot \hat{#2}}   $}}
\new{\Sc}{Schr\"{o}dinger}
\new{\sa}{self-adjoint}
\new{\se}{Schr\"odinger's equation}
\new{\BM}{Bohmian mechanics}
\new{\qf}{quantum formalism}
\new{\qm}{quantum mechanics}
\new{\wf}{wave function}
\new{\ewf}{effective wave function}
\new{\cwf}{conditional wave function}
\new{\oqm}{orthodox \qm}
\new{\qe}{quantum equilibrium}
\new{\rv}{random variable}
\new{\hv}{hidden variable}
 \new{\PV}{projection--valued--measure}
\new{\POV}{positive operator valued measure}
\new{\born}{\rho=|\psi|^2}
\new{\ai}{\alpha\in {\cal I}}
\renew{\a}{\alpha}
\new{\suma}{\sum_{\a \in I }}
\new{\ot}{\otimes}
\new{\bigo}{\bigoplus}
\new{\la}{\lambda_{\a}}
\new{\biga}{\bigoplus_{\a}}
\new{\psia}{\psi_{\a}}
\new{\Phia}{\Phi_{\a}}
\new{\Ha}{{\H}_{\a}}
\renew{\H}{\mbox{${\cal H}$}}
\new{\R}{\mbox{${\rm I\!R}$}}
\new{\F}{{\cal F }}
\new{\E}{\mbox{${\cal E}$}}
\new{\Ex}{\mbox{${\cal E}$}}
\renew{\P}{\mbox{${\rm I\!P}$}}
\new{\M}{\mbox{${\cal M}$}}
\new{\N}{\mbox{${\cal N}$}}
\new{\Pa}{P_{\a}}
\new{\LA}{\Lambda}
\renew{\O}{{O}}
\new{\As}{{R}}
\new{\Aa}{R_{\a}}
\new{\Aad}{R^{\dagger}_{\a}}
\new{\Al}{R_{\lambda}}
\new{\Ald}{R^{\dagger}_{\lambda}}
\new{\Ay}{R_y}
\new{\Ayd}{R^{\dagger}_{y}}
\new{\pr}{\mbox{Prob}\,}
\new{\prp}{\mbox{Prob}_\psi\,}
\new{\I}{\mbox{${\cal I}$}}
\new{\lam}{\lambda}
\new{\leta}{\lambda_{\beta}}
\new{\vpsi}{v^\psi }
\new{\vpsit}{v^{\psi _t}}
\new{\lit}{ \lim_{t \to\infty}}
\new{\id}{\mbox{\rm I}}

{\begin{list}{}%
{\settowidth{\labelwidth}{#1}%
\setlength{\leftmargin}{\labelwidth}%
\addtolength{\leftmargin}{\labelsep}%
\setlength{\parsep}{0pt}%
\setlength{\itemsep}{0pt}%
\setlength{\topsep}{0pt}%
}}%
{\end{list}}%
\begin{document}

\draft
%
\title{ A Survey of Bohmian Mechanics}
\author{K. Berndl, M. Daumer, and D. D\"{u}rr}
\address{Mathematisches Institut der Universit\"{a}t M\"{u}nchen,\\
Theresienstra{\ss}e 39,
80333 M\"{u}nchen, Germany}
\author{S. Goldstein}
\address{Department of Mathematics,
Rutgers University,\\ New Brunswick, NJ 08903, USA}
\author{N. Zangh\`{\i}}
\address{ Dipartimento di Fisica,
Universit\`a di Genova, Sezione INFN Genova,\\ Via Dodecaneso 33,
16146 Genova, Italy}

\maketitle
\thispagestyle{myheadings}
\markright{To appear in {\sc Il Nuovo Cimento}}
\begin{abstract}\BM\ is the most naively obvious embedding imaginable  of
\Sc's equation
into a completely coherent physical theory.  It describes a world in which
particles move in a highly non-Newtonian sort of way, one which may at
first appear to have little to do with the spectrum of predictions of
quantum mechanics.  It turns out, however, that as a consequence of the
defining dynamical equations of \BM, when a system has \wf\ $\psi$ its
configuration is typically random, with probability density $\rho$ given by
$|\psi|^2$, the \qe\ distribution.  It also turns out that the entire \qf,
operators as observables and all the rest, is a consequence of \BM.
\end{abstract}

\section{Bohmian Mechanics in a Nutshell}

 Suppose that when we talk about the \wf\ of a system of $N$
particles,
we seriously mean what our language conveys, i.e., suppose we insist that
{\sl ``particles'' means particles\/}. If so, then the \wf\ cannot provide a
complete description of the state of the system; we must also specify its
most important feature, the {\it positions\/} of the particles themselves!

 Suppose, in fact, that the complete description of the quantum
system---its {\it state}---is given by
$$ (Q \,,\,\psi)$$
where
  $Q=({\bf Q}_1   \dots    {\bf Q}_N) \in \R ^{3N},$ with
${\bf Q}_k$  the  positions  of particles, and
$\psi=\psi(q)=\psi ({\bf q}_1   \dots    {\bf q}_N) $ is the \wf.
Then we shall have a theory once we specify the law of  motion for the
state $(Q,\psi)$.  The simplest possibility is that this motion is given by
first-order equations---so that $(Q,\psi)$ is indeed the state in the sense
that its present specification determines the future.  We already have an
evolution equation for $\psi$, i.e., \Sc's equation,
\be\label{eqsc}i\hbar\frac{\partial \psi }{\partial t}  =
-{\sum}_{k=1}^{N}
\frac{{\hbar}^{2}}{2m_{k}}\Delta_{k}\psi  + V\psi \,.     \ee
According to what we have just said we are looking for an evolution
equation for $Q$ of the form
\be\label{eqvel}\frac {dQ}{dt}=\vpsi(Q) \ee
where $\vpsi =
({\bf v}^{\psi}_{1}   \dots  {\bf  v}^{\psi}_{N})$.
 Thus the role
of  $\psi$  is to choreograph a motion of particles  through the
vector field on configuration space that it defines,
$$ \psi \to \vpsi \,.$$

But how  should $\vpsi$ be chosen?
 A specific form for $v^\psi$ emerges
by requiring space-time symmetry---Galilean and time-reversal invariance
(or covariance), and ``simplicity''  \cite{DGZ92a}:

\noindent For one-particle system we find
$$
{\bf v}^{\psi} = \frac{\hbar}{m}{\rm Im}\frac{
\bold{\nabla}\psi}{\psi}, $$
and for a general $N-$particle system
\be\label{velo}
{\bf v}_{k}^{\psi} = \frac{\hbar}{m_{k}}{\rm
Im}\frac{ \bold{\nabla}_{k}\psi}{\psi}\,.
\ee

We've arrived at {\it Bohmian mechanics\/}, defined by (1--3) for a
nonrelativistic system (universe) of $N$ particles, without spin. This
theory, a refinement of de Broglie's pilot wave model, was found and
compellingly analyzed by David Bohm in 1952
\cite{Boh52a,Boh52b,Boh53,BH93,Bel87,DGZ92a,DGZ93b,DGZ92c,Hol93,Alb94}.
Spin, as well as Fermi
and Bose-Einstein statistics, can easily be dealt with
 and in fact arise in a natural
manner \cite{Bel66,Boh52a,Nel85,Gol87,DGZ94}.
\medskip

 Let us brief\/ly mention how to incorporate {\it spin} into \BM..  Note that
on the
right-hand side of the
equation for the velocity field  the  $\bold{\nabla}$
is suggested by rotation invariance, the $\psi$ in the denominator by
homogeneity, the ``${\rm Im}$'' by time-reversal invariance, and the
constant in  front is precisely what is required for covariance under Galilean
boosts. Rotation invariance requires in particular that rotations act on the
value space of the \wf. But the latter action is rather inconspicuous for
spinless
particles. The simplest nontrivial (projective) representation of the rotation
group is the $2$-dimensional ``spin $\frac{1}{2}$'' representation. This
representation leads to a \BM\ involving spinor-valued \wf s for a single
particle (and spinor-tensor-product-valued \wf\ for many particles).
Beyond the fact that the wave function now has a more abstract value space,
nothing much changes from our previous description:
The \wf\ evolves according to a Hamiltonian that contains the Pauli term,
for a single particle  proportional to ${\bf B \cdot}{\bold{\sigma}} $,
which
represents the coupling between the
``spin'' and an external magnetic field $\bf B$.  The
configuration evolves according to the natural extension of the velocity field
to spinors, obtained, say, by multiplying both the
numerator and denominator of the argument of ``Im'' on the left by $\psi^*$
And interpreting the result for the case of spinor values as
a spinor-inner-product:
$$
{\bf v}^{\psi} = \frac{\hbar}{m}{\rm Im}\frac{\psi^* \bold{\nabla}\psi}
{\psi ^*\psi}.
$$
\medskip

A remark on {\it Bose-Fermi statistics}: According to orthodox quantum
mechanics, the very notion of {\it indistinguishable particles} seems to be
grounded on the nonexistence of particle trajectories and on the practical
impossibility of distinguishing identical particles at two different times.
This might lead to the expectation that it should be quite problematical
to incorporate the description of indistinguishable particles into \BM.
However, this is not so. Indeed, the usual symmetry conditions on the \wf\
arise naturally when the Bohmian approach is applied to systems of
indistinguishable particles. Moreover, when spin is taken into account, the
fact that the intermediate statistics (the so called {\it parastatistics})
are to be excluded turns out to be a {\it consequence of the very existence
of trajectories} (as does the fact that in a two dimensional world there
would be many more possibilities than just bosons and fermions)
\cite{DGZ94}.
\bigskip

\BM\ is a fully deterministic theory of particles in motion, but a
motion of a profoundly nonclassical, non-Newtonian sort.  We should remark,
however, that in the limit $\frac{\hbar}m\rightarrow 0$, the Bohm motion
$Q_t$
approaches the classical motion.

 But what does this theory, \BM, have to do with orthodox quantum theory,
i.e., with the \qf? Well, of course, they share \Sc's equation.  However, in
orthodox quantum theory noncommuting observables, represented
by
self-adjoint operators, play a fundamental role, while they do not appear
at all in the {\it formulation\/} of \BM. Nonetheless, it can be shown that
\BM\ not only accounts for quantum phenomena---this was essentially done
by Bohm  in 1952 and 1953---but also embodies
the quantum formalism itself, self-adjoint operators, randomness given by
$\rho=|\psi|^2$, and all the rest, as the very expression of its empirical
import \cite{DGZ92a,DDGZ94}.
\bigskip

Equations \eq{eqvel} (together with \eq{velo}) and \eq{eqsc} form a
complete specification of the theory. There is no need, and indeed no room,
for any further axioms. As for the status of the the familiar distribution
$\rho =| \psi|^2$ in \BM, an answer is provided by reflecting upon the role
of equilibrium measures for dynamical systems.  Suppose one is interested
in aspects of, say, the long time behavior, of patterns of statistical
regularities which occur. Then some of the most basic of such information
is usually provided by a measure stationary for the dynamics, so finding
such a measure is often the key step in the analysis.  Now it turns out
that for
\BM\ there is, in fact, no useful stationary measure, since the velocity
field is typically time-dependent. Yet, $| \psi|^2$ is as good as a
stationary measure. This distribution is in fact {\it equivariant\/}:

Consider an arbitrary initial ensemble $\rho$  and let
$$\rho \to \rho_t$$
be the  ensemble evolution  arising from  Bohmian motion. If
 $\rho= \rho^\psi$  is a functional of $\psi$ we may also consider  the
ensemble evolutions arising from \se\
$$\rho^\psi \to \rho^{\psi_t}\,.$$
 $\rho^\psi$ is {\it equivariant  \/}  if these
evolution are compatible  $$\big(\rho^\psi\big)_t =
\rho^{\psi_t}$$
 That $\rho = |\psi|^2$ is
equivariant follows from comparing the quantum f\/lux equation
\be\label{flux}
 \frac{\partial |\psi|^2}{\partial t} + \hbox{\rm div}\, J^\psi =0
\ee
where $J^\psi = ({\bf J}_1^\psi  \dots  {\bf J}_N^\psi )\, $, $\;
{\bf J}_k^\psi= \frac{\hbar}{m_k}{\rm Im}\ (\psi^*
\nabla_k\psi)$,
with  the continuity equation associated with particle motion
$$\frac{\partial \rho}{\partial t} + \hbox{\rm div}\ \big( \rho \vpsi\big) =0
$$
\smallskip
Since $J^\psi=  v^{\psi}\, |\psi|^2 $, the continuity equation is satisfied for
$\rho=
|\psi|^2$. Thus:

\bqu
{\it If $\rho(q, t_0) = |\psi(q, t_0)|^2 $ at some time $t_0$  then
$\rho(q, t) = |\psi(q, t)|^2$ for all $t$.  }\equ

 Suppose now that a system has \wf\ $\psi$. We shall call the probability
distribution on configuration space given by $\rho=|\psi|^2$ the {\it
quantum equilibrium} distribution. And we shall say that a system is in
quantum
equilibrium when its configuration are randomly distributed according to the
quantum equilibrium distribution.  The empirical implications of \BM\ are
based on the following

\bqu{{\it Quantum equilibrium hypothesis (QEH): When a system has \wf\
$\psi$, the distribution $\rho$ of its
configuration satisfies $\;\rho = |\psi|^2$}}. \equ

\section{Existence of Quantum Trajectories}

 Before proceeding to a sketch of how \BM\ accounts for quantum
phenomena,  we shall address the problem of whether \BM\ is
a mathematically sound theory.  After all, the velocity field \eq{velo} reveals
rather obviously possible catastrophic events for the motion: $\vpsi$ is
singular at the {\it nodes} of $\psi$, i.e.,at points where $\psi=0$.  We shall
consider then the   defining equations of   {\it \BM}
\ba
  \frac {dQ}{dt}\! &=&\! \vpsi(Q) \nonumber  \\
  i\hbar\frac{\partial \psi }{\partial t}\! &=&\! H \psi, \nonumber
\ea
where $\vpsi$ is given by (3) and $H\psi$ is the right hand side of (1), and
inquire about {\it the existence and uniqueness} of their solutions.

The ``problem of the existence of dynamics'' for \Sc 's equation is usually
reduced to showing that the
relevant Hamiltonian $H$ (given by the
particular choice of the potential $V$) is self-adjoint. This has been done in
great
generality, independent of the number of particles and for large classes of
potentials, including singular potentials like the Coulomb potential, which is
of
primary physical interest \cite{Kat51,RS75}.
In \BM\ we have not only \Sc's equation  to consider but also the
differential
equation  governing the motion of the particles. Thus the
question of {\it existence of the dynamics of \BM\/}  depends now on
detailed
regularity properties of the velocity field $\vpsi$.
 {\it Local} existence and uniqueness of Bohmian trajectories
is guaranteed if the velocity field $\vpsi$ is locally Lipschitz continuous. We
therefore certainly need greater regularity for  the \wf\ $\psi$ than
merely that $\psi$ be in $L^2$.
{\it Global} existence is more delicate: In addition to the nodes of $\psi$,
there are
singularities  comparable to those of Newtonian mechanics. Firstly,

even
for a globally smooth velocity field the solution $Q_t$ may explode,
i.e., it may reach infinity in finite time. Secondly, the singular points of
the potential, are ref\/lected in
singular behavior of the \wf\ at such points, giving rise to singularities in
the
velocity field. (For example, the ground state \wf\
of one particle in a Coulomb potential $V(q)=1/|q|$, $q\in \R ^3$
(``hydrogen atom'') has the form $e^{-|q|}$, which is not
differentiable at the point $q=0$ of the potential singularity.)

The problem  is then the  following: Suppose that at some
arbitrary ``initial time'' ($t_0=0$) the $N$-particle configuration lies in the
complement of the set of nodes and singularities of $\psi _0$. Does  the
trajectory develop in a finite amount of time into a singularity of the
velocity
field $\vpsi$, or does it reach infinity in finite time?
It turns out that  the
answer is negative for ``typical'' initial values and a large class of
potentials,
including the physically most interesting case of $N$-particle Coulomb
interaction with arbitrary charges and masses. Our results
\cite{BDGPZ93,BDGZ93,BDGPZ94} are summarized by the
following

\begin{prop} For a large class of Hamiltonians (including
Coulomb with arbitrary charges and masses) and sufficiently regular initial
datum $\psi_0$ the solution exists uniquely and  globally in time for
$| \psi_0|^2$-almost all initial configurations $Q_0$. \end{prop}

The quantity of central importance for our proof \cite{BDGPZ94} of this
theorem turns out
to be the quantum current
$j^{\psi} = (J^{\psi} ,|\psi |^2)$, with $J^{\psi} =
\vpsi |\psi|^{2}$ the quantum probability f\/lux. The absolute value of the
f\/lux through any surface in configuration-space-time
controls the probability that a trajectory crosses that surface. Consider
a smooth surface $\Sigma$ in configuration-space-time. The expected number
of crossings of $\Sigma$ by the random trajectory $ Q_t$ is given by
\[ \int _{\Sigma}|j_t(q) \cdot n| d\sigma  \]
where $n$ denotes the local unit normal vector at $(q,t)$.
($\int_{\Sigma}(j \cdot n)d\sigma $ is the expected number of {\em
signed}\/ crossings.)  To get a handle on this consider first a small
surface element which the trajectories cross at most once. The density of
crossings is readily calculated to be $|j \cdot n|$.  Invoking the
linearity of the expectation value yields then the general statement. (In
this regard we note that for the related problem in stochastic mechanics
\cite{Nel85} the particle trajectories are realizations of a diffusion
process and are hence not differentiable, i.e., velocities do not exist.
Thus in stochastic mechanics the current does not have the same probabilistic
significance and our analysis does not apply to stochastic mechanics.)
Surfaces relevant to our analysis are those formed by the boundaries of
neighborhoods around all the singular points for \BM. Loosely
speaking, the importance of the quantum f\/lux is grounded in the insight:
``If there is no absolute f\/lux into the singular points, the singular
points are not reached.''

\section{Empirical Implications}

A  systematic analysis of the empirical implications of \BM\ falls
naturally into two parts: \\
({\bf A})  The emergence and significance of other (nonconfigurational)
observables.
\\ ({\bf B}) The clarification and justification of the QEH.

As for ({\bf B}), compare  the QEH  with the Gibbs postulate (GP) of
statistical
mechanics:
$$\mbox{\it quantum equilibrium\/}\qquad \rho = |\psi|^2$$
$$\mbox{\it thermodynamic equilibrium\/}\qquad
\rho\sim e^{-\beta H} $$
While  the complete justification of the GP is remarkably difficult (and as of
now is
nonexistent), that of the QEH is relatively easy \cite{DGZ92a}.

 As for ({\bf A}), the crucial observation has been made by  Bell \cite{Bel82}:
\bqu  {\dots    in physics the only observations we must consider are
position
observations,  if only the positions of instrument pointers.   It is a great
merit of the de Broglie-Bohm picture to force us to consider this fact.   If
you
make axioms  rather than definitions and theorems  about the
`measurement' of anything else  then you commit redundancy and risk
inconsistency.}  \equ

When one comes to ``measurements'' and ``observables,'' a  warning against
the misuse of these words is mandatory. We  again quote Bell \cite{Bel90}:
\bqu{ \dots The first charge against ``measurement,'' in the fundamental
axioms of
quantum mechanics, is that it anchors the shifty split of the world into
``system'' and ``apparatus.'' A second charge is that the word comes loaded
with meaning from everyday life, meaning which is entirely inappropriate in
the quantum context. When it is said that something is ``measured'' it is
difficult not to think of the result as referring to some {\it preexisting
property\/} of the object in question. This is to disregard Bohr's insistence
that in quantum phenomena the apparatus as well as the system is
essentially involved.  \dots Even in a lowbrow practical account, I think it
would be good to replace the word ``measurement,'' in the formulation, with
the word ``experiment.'' }\equ

\subsection{Experiments}
 When we speak of a very general experiment \E, beginning, say, at
$t=0$
and ending at time $T$,  we have in mind a fairly
definite initial   state $\Phi_0 =\Phi_0(y)$ of the apparatus, one for which
the apparatus
should function   as intended, as well as a definite initial state of the
system
$\psi=\psi(x)$ on which the experiment is performed. Under these conditions
it turns out that
 the composite system formed by system and apparatus, with generic
configuration $q=(x,y)$, has initial
\wf:
$$
\Psi_0 = \psi \ot \Phi_0 .$$
Moreover, $\E$ will be specified by a
unitary operator $U$ generating the time evolution arising from the
interaction of the system and apparatus,   which yields the \wf\ $\Psi_T$ of
the composite system after the experiment; and a calibration function $F$
from the configuration space of the composite system to some value space,
e.g. $\R$, fixing the {\it scale}   of the experiment, and
defining the result ${\sf Z}\equiv F(Q_T)$ of the experiment---think of the
``orientation of the apparatus pointer'' or some coarse-graining
thereof---as a function of the configuration $Q_T$ of the system and
apparatus
after the experiment.
\bigskip

 Assume QEH. Then   $Q_T$ is randomly distributed
according to the quantum equilibrium measure $\P_T(dq)=|\Psi_T|^2dq$
and   ${\sf Z}$ is a random variable (on the probability space of the initial
configurations of system and apparatus) with distribution given
by the probability measure
$$ \mu  =\P_T\circ F^{-1}. $$

A stepping stone of our analysis \cite{DDGZ94} is the following
\begin{prop} With any experiment \E\ there is always associated a
positive-operator-valued measure (POV) $O(d\lam)$ such that
$$ \mu(d\lam) = \langle\,\psi\,, O(d\lam)\,\psi\rangle$$ \end{prop}

This follows very easily from the observation that the map $\psi \to \mu$
from (initial system) \wf s to probability measures on the value space,
explicitly given
by the following sequence of maps
$$\psi \to \Psi = \psi \otimes \Phi_0 \to \Psi_T \to \P_T \to \mu= \P_T
\circ F^{-1} \,,$$
is  a  normalized bilinear map on the system Hilbert space \H,  since
the
middle map to the \qe\ distribution,
$$\Psi_T \to \P_T(dq) = \Psi_T^{*} \Psi_T dq ,$$  is obviously
bilinear, while all the other maps are linear, all but the second trivially so.
Now, by elementary functional analysis, the notion of such a bilinear map is
completely equivalent to that of a POV!   We note that when the experiment
is ``measurement-like'' (by this we merely mean that, unlike a coin f\/lip,
the
outcome is {\it reproducible}) the POV $O$ is actually a {\it projection valued
measure} (PV) and with every such experiment we may associate a
self-adjoint operator $A$,
$$ \E \to A,$$
which governs the statistics of the outcomes in the usual way \cite{DDGZ94}.

We recall that because of difficulties in the application of the usual operator
formalism,  it has been proposed in the framework of the so called {\it
operational approach} to \qm\ that we should go beyond
operators-as-observables, to ``{\it generalized observables\/}''
\cite{Dav76,Hol82,Kra83,Lud83a}. The basis of
this extension lies in the observation that, by the
spectral theorem, the concept of self-adjoint operator  is completely
equivalent to that of (normalized) projection-valued measure (PV)
on the value space $\R$.  Since orthogonal projections are
among the simplest examples of positive operators, a natural generalization
of a ``quantum observable'' is then provided by a (normalized) {\it
positive-operator-valued measure} (POV)---when a POV  is sandwiched by
a \wf\ it generates a probability distribution.

On the other hand, the emergence and role of POV's in \BM\ is not  a matter
of generalization; rather it is merely an expression of
the bilinearity of \qe\ together with the linearity of \Sc's evolution. Thus
the
fact that with {\it every} experiment is
associated a POV, which forms a compact expression of the statistics for the
possible results, is a near mathematical triviality. It is therefore rather
dubious that the occurrence of POV's as observables---the simplest case of
which is that of PV's---can be regarded as suggesting any deep truths about
reality or about epistemology.  In particular,  so understood, the notion of
\sa-operator-as-observable $A$ in no way
implies that anything is really being measured in the experiment with which
$A$ is associated, and
certainly not the operator $A$ itself!  In a general experiment no property is
being measured, even if the experiment happens to be
{\it measurement-like}.
(In this regard  we note that experiments associated with the position
operator are for the most part  an important exception, though there are
``measurements'' of the position operator that are not measurements of the
{\it actual} position \cite{ESSW92,DFGZ93f,DHS93,DDGZ94}.)
 \medskip

 That  \sa\ operators are associated only with {\it special}
experiments is a
further indication that  the usual \qf, based only on  \sa\ operators, is
merely an idealization, rarely directly relevant in practice. Indeed, a great
many significant real-world experiments are simply not at all associated
with operators in the usual way \cite{DG93,DDGZ93,DDGZ94,Lea93}.

Consider for example an electron with fairly general initial \wf, and
surround the electron with a ``photographic'' plate, away from (the support
of the \wf\  of) the electron, but not too far away. This set-up measures the
position of ``escape'' of the electron from the region surrounded by the plate.
Notice that since in general there is no definite time of escape, it is not at
all
clear which operator should correspond to the escape position. Indeed, it can
be shown \cite{DDGZ93,DDGZ94}
that there is no such operator, that for the
experiment just described the probabilities for the possible results cannot be
expressed in  the usual form, and in fact are not given by the spectral
measure for any operator.

We note that the study of the asymptotic limit for this
situation---the scattering regime---is the
starting point for a reformulation of scattering theory \cite{DDGZ93} based
on the so called
{\it scattering-into-cones-theorem}, proved by Dollard \cite{Dol69},
and the {\it f\/lux-across-surfaces-theorem}  \cite{CNS75}, of which a
complete proof is still lacking.
\bigskip

\subsection{Quantum Equilibrium}\label{secPQE}
 We'd like now to turn to the clarification and
justification of the QEH \cite{DGZ92a,DGZ92c,DGZ93b}.
There are some crucial subtleties in the QEH, which we can begin to
appreciate by first asking the question: Which systems should be governed
by \BM? The systems which we normally consider are subsystems of a
larger
system---for example, the universe---whose behavior (the behavior of the
whole) determines the behavior of its subsystems (the behavior of the
parts). Thus for a Bohmian universe, it is only the universe itself which a
priori---i.e., without further analysis---can be said to be governed by
\BM. So let's consider such a universe. Our first difficulty
immediately emerges: In practice $\born$ is applied to (small) subsystems.
But only the universe has been assigned a \wf\ (which we shall now denote
by $\Psi$)! What is meant then by the RHS of $\born$, i.e., by the \wf\ of a
subsystem?

 Let's go further. Fix an {\it initial} \wf\ $\Psi_0$ for this universe. Then
since the Bohmian evolution is completely deterministic, once the {\it initial}
configuration $Q$ of this universe is also specified, all future events,
including of course the results of measurements, are determined. Now let
$X$ be some subsystem variable---say the configuration of the subsystem at
some time $t$---which we would like to be governed by $\born$. But how
can
this possibly be, when there is nothing at all random about $X$?

Of course, if we allow the {\it initial} universal configuration $Q$ to be
random, distributed according to the \qe\ distribution ${|\Psi_0(Q)|}^2$,
it follows from equivariance that the universal configuration $Q_t$ at
later times will also be random, with distribution given by ${|\Psi_t|}^2$,
from which you might well imagine that it follows that any variable of
interest, e.g., $X$, has the ``right'' distribution. But even if this is so
(and it is), it would be devoid of physical significance! What possible
physical significance can be assigned to an ensemble of universes, when we
have but one universe at our disposal, the one in which we happen to
reside? We cannot perform the {\it very same\/} experiment more than
once.
But we can perform many similar experiments, differing, however, at the
very least, by location or time. In other words, insofar as the use of
probability in physics is concerned, what is relevant is not sampling
across an ensemble of universes, but sampling across space and time within
a single universe.  What is relevant is empirical distributions---actual
relative frequencies for an ensemble of actual events.

Two problems must thus be addressed, that of the meaning of the
\wf\ $\psi$ of a subsystem and that of randomness. It turns out that once
we come to grips with the first problem, the question of randomness almost
answers itself. We obtain just what we want---that $\born$ in the sense of
empirical distributions; we find that in a {\it typical\/} Bohmian universe an
appearance of randomness emerges, precisely as described by the \qf.

What about the \wf\ of a subsystem? Given a subsystem we may write
$q=(x,y)$ where $x$ and $y$ are generic variables for the configurations of
the subsystem and its environment. Similarly, we have $Q_t=(X,Y)$ for the
actual configurations (at time $t$). What is the simplest possibility for
the wave function of the subsystem, the $x$-system; what is the simplest
function of $x$ which can sensibly be constructed from the actual state of
the universe at time $t$ (which we remind you is given by $Q_t$ and
$\Psi_t=\Psi$)? Clearly the answer is what we call the {\it \cwf}
$$
\psi(x)=\Psi(x,Y).
$$
This is all we need! (This is not quite the right
notion for the ``effective'' \wf\ of a subsystem, upon which we shall
elaborate in the next
section, but whenever the latter exists
it agrees with what we have just described.) Now see what you can do
without
actual configurations! (You'll, of course, quickly encounter the
measurement problem!)
\bigskip

The main result of our analysis \cite{DGZ92a} is summarized by the following
\begin{prop}
When a system has wave function $\psi$,
the distribution $\rho$ of its configuration typically satisfies $\;\rho =
|\psi|^2$. \end{prop}

This means that for {\it typical} initial configurations of the universe, the
empirical distribution of  an ensemble of $M$ identical subsystems with
\wf\
$\psi$ converges to $\;\rho = |\psi|^2$ for large $M$.
The statement refers to an equal-time ensemble or to a multi-time ensemble
and the notion of typicality
is expressed by the measure $ \P^{\Psi_0}(dQ)$ and more importantly by the
conditional measure $\P^{\Psi_0}(dQ| \M)$, where the set \M\ takes into
account any kind of {\it prior} information---always present---ref\/lecting
the
macroscopic state at a time prior to all experiments. Moreover, the above
proposition holds
under physically minimal conditions, expressed by certain measurability
conditions ref\/lecting the requirement
that {\it facts} about results and initial experimental conditions not be
forgotten.

\section{The Effective Wave Function}

Let's pause for a moment and get familiar with the
notion of
conditional \wf\ by looking at a very simple example:

 Consider two particles in one dimension,  whose
evolution is
governed by the Hamiltonian
$$
H =H^{(x)}+H^{(y)} +H^{(xy)} = -\frac{\hbar^2}{2m}\big(
\frac{\partial^2}{\partial x^2} +
\frac{\partial^2}{\partial y^2}\big) \ + \frac{1}{2} \kappa(x-y)^2 .
$$
For simplicity let us set $\hbar=m=\kappa=1$. Assume that the composite
has
initial \wf
$$
\Psi_0 = \psi \ot \Phi_0 $$
$$\hbox{with}\quad \psi(x)=\pi^{-\frac{1}{4}}
 e^{-\frac{x^2}{2}}\quad\hbox{and}\quad
\Phi_0 (y)=\pi^{-\frac{1}{4}} e^{-\frac{y^2}{2}}.
$$
By solving the basic equations of \BM\    one easily obtains that
$$
\Psi_t (x,y) =\pi^{-\frac{1}{2}} (1+it)^{-\frac{1}{2}}
 e^{-\frac{1}{4}\big[(x- y)^2+\frac{(x+y)^2}{1+2it}\big]}, $$ and $$X_t =
a(t)X + b(t)Y \quad\hbox{and}\quad Y_t = b(t)X +a(t) Y , $$ where $a(t)=
\frac{1}{2}[ (1+t^2)^{\frac{1}{2}}+1] $, $b(t)= \frac{1}{2}[
(1+t^2)^{\frac{1}{2}}-1] $, and $X,Y$ are the initial conditions of the two
particles.  Focus now on one of the two particles (the $x$-system) and
regard the other one as its environment (the $y$-system).  The conditional
\wf\ of the $x$-system $$\psi_t(x) = \Psi_t (x, Y_t) $$ depends, through
$Y_t$, on {\it both} the initial conditions for the environment {\it and}
the initial condition for the particle. In other words, the evolution of
$\psi_t$ is random, with probability law determined by $|\Psi_0|^2$. In
particular, $\psi_t$ does not satisfy \Sc's equation for any $H^{(x)}$.
\bigskip

We remark that even when the $x$-system is dynamically decoupled from its
environment, the conditional \wf\ will not in general evolve according to
\Sc's equation. Thus the conditional \wf\ lacks the {\it dynamical}
implications from which the \wf\ of a system derives much of its physical
significance. These are, however, captured by the notion of \ewf:

 Suppose that
\be\label{starstar}
\Psi(x,y)=\psi(x)\Phi(y)+\Psi^\perp(x,y) \,,
\ee
where $\Phi$ and $\Psi^\perp$ have macroscopically disjoint $y$-supports.
If
$$
Y\in \mbox{supp}\,{\Phi}
$$
we say that $\psi$ is the {\it \ewf\/} of the $x$-system. Of course, $\psi$
is also the \cwf---nonvanishing scalar multiples of
\wf s are  naturally identified. (In  fact, in \BM\ the \wf\  is naturally a
projective object
since \wf s differing by a multiplicative
constant---possibly time-dependent---are associated with the same
vector field, and thus  generate the same dynamics.)

One might wonder why systems ever possess an effective wave function.
In fact, in general they don't!  For example the $x$-system will not have an
\ewf\
when, for example, it belongs to a larger microscopic system whose \ewf\
doesn't factorize in the appropriate way.
However, the {\it larger} the environment of the $x$-system , the {\it
greater} is the potential for the existence of an \ewf\ for this system, owing
in effect to the abundance of ``measurement-like'' interactions with a larger
environment. The notion of \ewf\ is {\it robust\/}, as there is a
natural tendency toward the formation of  stable  \ewf s  via {\it
dissipation}:
Suppose that initially the $y$-supports of $\Phi$ and
$\Psi^{\perp}$ are just ``sufficiently'' (but not macroscopically) disjoint;
then,
due to the interactions with the environment, the amount of $y$-disjointness
will tend to increase dramatically as time goes on, with, as in a chain
reaction, more and more degrees of freedom participating in this
disjointness. When the effect of this dissipation, or ``decoherence,'' are
taken
into account, one find that even a small amount of $y$-disjointness will often
tend to become ``sufficient,'' and quickly ``more than sufficient,'' and
finally
macroscopic.

The ever-decreasing possibility of interference between macroscopically
distinct \wf s due to typically uncontrollable interactions
with the environment is nowadays often referred to as {\it decoherence}
(Griffiths \cite{Gri84}, Omnes \cite{Omn88}, Leggett \cite{Leg80}, Zurek
\cite{Zur82}, Joos-Zeh \cite{JZ85}) and has been
regarded (Gell-Mann-Hartle \cite{GMH90}) as a crucial ingredient for
extracting a ``quasiclassical domain of familiar experience'' from the
quantum formalism itself (see also \cite{GP94}).  One of the best
descriptions of the mechanism of decoherence, though not the word itself,
can be found in the Bohm's 1952 ``hidden variables'' paper\cite{Boh52a}.
We wish to emphasize, however, as did Bell in his article ``Against
Measurement''
\cite{Bel90}, that decoherence in no way comes to grips with the measurement
problem itself, being merely a {\it necessary}, but not a sufficient,
condition for its complete resolution.  In contrast, the very notion
of \ewf\ resolves the measurement problem at once.

Consider for example an experiment \E\ with an apparatus so  designed that
there are  only  finitely (or countably) many  possible {\it outcomes},
labeled by $\alpha\in I$.
Then, after the
experiment the \wf\ of the composite is of the form
\be\label{ormfin} \Psi_T = \suma \psia \ot \Phia, \ee
where the $\Phia$ are (normalized) apparatus states supported by the
macroscopically distinct sets $\a\in I$ of apparatus  configurations.
Of course, for \BM , the terms of \eq{ormfin}  are not all on the same
footing: one of them, and only one, is selected, or more precisely supported,
by the  outcome---corresponding, say, to $\a_0
$---which {\it actually} occurs.
 It follows that after the experiment, at time $T$, the $x$-system has
\ewf\ $\psi_{\a_0}$. This is how {\it collapse} (or {\it reduction}) of the
\ewf\ to the one associated with the outcome $\a_0$ arises in \BM.
\bigskip

Note that while in orthodox quantum theory the collapse is
merely superimposed upon the unitary evolution---without a precise
specification of the circumstances under which it may legitimately be
invoked---we have now, in \BM , that the evolution  of the \ewf\ is
actually given by
a stochastic process, which consistently embodies {\it both} unitarity {\it
and}
collapse as appropriate. In particular, the \ewf\ of a subsystem evolves
according to \se\ when this system is suitably isolated. Otherwise it ``pops in
and out'' of existence in a random fashion,  in a way determined by  the
continuous (but still random) evolution $\psi_t$ of the conditional \wf.
(In this regard, as far as the general problem of chaotic behavior in quantum
theory
is concerned, note that there is nothing in \BM\ which would preclude sensitive
dependence
on initial conditions, of $Q_t$ on $Q_0$ and $\psi_0$, and hence positive
Lyapunov exponents. In \BM\ ``quantum chaos'' arises, as in the classical
case,
 solely from the dynamical law and not  from
the collapse rule applied in measurements \cite{DGZ92b}.)
\bigskip

\section{Quantum Physics without Quantum Philosophy}

 We would like to make a few comments now about \BM\ and
``the real world.'' There is at best an uneasy truce between orthodox
quantum theory and the view that there is an objective reality, of a more
or less familiar sort on the macroscopic level.  Recall, for example, \Sc's
cat. What does \BM\ contribute here? In a word, everything! A world of
objects, of large collections of particles which combine and move more or
less as a whole, presents no conceptual difficulty for \BM, since \BM\ is
after all a theory of particles in motion and allows for the {\it
possibility\/} of such large collections.

So what, when all is said and done, does the incorporation of the particle
positions, of the configurations, buy us? A great deal:
\begin{enumerate}
\item randomness
\item familiar (macroscopic) reality
\item the \wf\ of a (sub)system
\item collapse of the wave packet
\item absolute uncertainty
\end{enumerate}
We have not yet explicitly addressed item  5.
 5 is a consequence of the analysis of
$\born$. It expresses the impossibility of obtaining
information about positions more detailed than what is given by the \qe\
distribution. It provides a precise, sharp foundation for the uncertainty
principle, and is itself an expression of global \qe \cite{DGZ92a}.
\bigskip

When all is said and done, \BM\ emerges as a
precise and coherent ``quantum theory''  providing a microscopic
foundation for the quantum formalism. To sum up,
 it seems fair to say that Bohmian
mechanics is nothing but quantum physics without quantum philosophy.
Moreover, the only objections which  are usually  raised against
\BM\ are  merely philosophical. Now we don't wish to enter here into
philosophical
disputes. We would, however,  like to mention that in response to the outrage
sometimes expressed towards the suggestion that particles might have
positions when they are not, or cannot be, observed, Bell, referring to
theories such as Bohm's, has said that
\bqu{ Absurdly, such theories are known as ``hidden variable'' theories.
Absurdly, for there it is not in the wave function that one finds an image
of the visible world, and the results of experiments, but in the
complementary ``hidden''(!) variables. Of course the extra variables are
not confined to the visible ``macroscopic'' scale.  For no sharp definition
of such a scale could be made. The ``microscopic'' aspect of the
complementary variables is indeed hidden from us.  But to admit things not
visible to the gross creatures that we are is, in my opinion, to show a
decent humility, and not just a lamentable addiction to metaphysics
\cite{Bel87a}. }\equ

\acknowledgements
This work was supported  by the DFG, by
NSF Grant  No. DMS-9305930, and by INFN.

\nocite{WZ83} \nocite{Mil90} \nocite{KEMN90}

\end{document}